\documentstyle[prl,multicol,aps,psfig]{revtex}
\def \beq{\begin{equation}}
\def \eeq{\end{equation}}
\begin{document}
\draft
\title{Fermions obstruct dimensional reduction in hot QCD}
\author{R.\ V.\ Gavai \cite{ervg} and Sourendu Gupta \cite{esg}}
\address{Department of Theoretical Physics, Tata Institute of Fundamental
         Research,\\ Homi Bhabha Road, Bombay 400005, India.}
\maketitle
\begin{abstract}
We have studied, for the first time, screening masses obtained from
glueball-like correlators in Quantum Chromodynamics with four light
dynamical flavours of quarks in the temperature range $1.5T_c\le
T\le3T_c$, where $T_c$ is the temperature at which the chiral transition
occurs.  We have also studied pion-like and sigma-like screening masses,
and found that they are degenerate in the entire range of $T$.  These
obstruct perturbative dimensional reduction since the lowest glueball
screening mass is heavier than them. Extrapolation of our results
suggests that this obstruction may affect the entire range
of temperature expected to be reached even at the Large Hadron Collider.
\end{abstract}
\pacs{11.15.Ha, 12.38.Mh\hfill TIFR/TH/00-17, hep-lat/0004011}

\begin{multicols}{2}
The quark-gluon plasma phase of QCD is currently of great experimental
interest. The recent announcement that the CERN heavy ion program may
have seen this phase, and the forthcoming start of the BNL Relativistic
Heavy Ion Collider are just two reasons for this interest.  The phase
structure of baryon-free QCD matter is pretty well determined through
lattice simulations \cite{lat}. While this was sufficient in the past,
experiments will soon begin to demand much more detailed information on
the high temperature phase. In this letter we report work that bears on
the detailed structure of this phase.

Lattice simulations of finite temperature ($T$) equilibrium field
theories use a discretisation of the Euclidean formulation for partition
function---
\begin{equation}
   Z(T)\;=\;\int{\cal D}\phi\exp\left[-\int_0^{1/T} dt\int d^3x
      {\cal L}(\phi)\right],
\label{part}\end{equation}
where $\phi$ is a generic field, $\cal L$ the Lagrangian density,
and the Euclidean ``time'' runs from 0 to $1/T$. The path integral
is over Bosonic (Fermionic) field configurations which are periodic
(anti-periodic) in Euclidean time.
For the lattice problem, $Z(T)$ is the trace of an appropriate power
of the transfer matrix, $\cal T$, in one of the spatial directions. 
All the information in the theory is encoded in the set of eigenvalues
of $\cal T$, $\Lambda_i(T)$.

Thermodynamics is determined by the largest eigenvalue, $\Lambda_0$, and its
derivatives with respect to various parameters of the
theory.  The spectrum of screening masses (inverse correlation lengths),
\begin{equation}
   \mu_i(T) \;=\; \log\left[\frac{\Lambda_0(T)}{\Lambda_i(T)}\right],
\label{scrm}\end{equation}
is in one-to-one correspondence with the remaining eigenvalues, and hence
contain all the remaining information about the $T>0$ physics of the theory.
Due to the lack of symmetry between the space and Euclidean time
directions in eq.\ (\ref{part}), $\cal T$ has only a subgroup of the
rotational symmetry of the $T=0$ Euclidean theory. Hence the spectrum
of $\mu_i$ is classified by the irreducible representations (irreps)
of this new symmetry group. These symmetries and their irreps have been
studied in \cite{glue,fermit}.

Dimensional reduction has provided a good qualitative picture of the high
temperature limit of field theories. This reduction yields an effective
theory describing equilibrium physics of the (3+1)-dimensional theory
at distance scales much greater than $1/T$.  At these distances only
fields with low momentum ($<T$) excitations survive; field excitations
with momentum scales of order $T$ or more are integrated out. In practice
this procedure is carried out perturbatively \cite{dimred}.  Since the
Fourier modes of Boson fields have momenta $2n\pi T$ in the Euclidean
time direction and Fermion modes have momenta $(2n+1)\pi T$ (where $n$
is a non-negative integer), therefore all Fermion modes and all non-zero
gauge field modes are integrated out.  The result of this integration
is a 3-dimensional theory of gauge fields coupled to adjoint scalars.

All the screening masses remaining at long distances should then belong to
the glue sector of the (3+1)-dimensional theory, with Fermions appearing
only in loop corrections.  This approach has had great success in
dealing with the electroweak sector of the standard model \cite{keijo},
and has been proposed as a way of dealing with high temperature QCD
\cite{braaten,yaffe}. A key question is about the temperature range
where this can be done reliably.

Lattice measurements test dimensional reduction in three ways---
\begin{enumerate}
\item
    The degeneracies of $\mu_i$ test whether the effective symmetries
    at high temperatures are those of a theory in lower dimension. This
    form of dimensional reduction has been seen in (3+1)-dimensional
    pure gauge $SU(2)$ and $SU(3)$ theory at temperatures as low as $2T_c$
    \cite{glue,saumen}.
\item
    Perturbative matching of the couplings of the dimensionally reduced
    theory can be tested by comparing its correlators and screening masses
    with those of the full theory. Comparison of the results of \cite{glue,saumen}
    with those in 3-d SU(2) \cite{owe} and SU(3) \cite{owe2} gauge
    theories with an adjoint scalar demonstrate such dimensional reduction,
    as do the results of \cite{karsch}. More recently, this reduction has
    also been tested in the gauge sector of $(2+1)$-dimensional QCD \cite{bengt}.
\item
    All low-lying screening masses must be shown to arise in the gauge
    sector of the theory if the perturbative matching is correct. Here we
    report that $(3+1)$-dimensional QCD with 4 degenerate flavours of
    light dynamical quarks fails this test.
\end{enumerate}

We have performed Hybrid Monte Carlo simulations of $SU(3)$
gauge theory with 4 flavours of dynamical staggered quarks at
temperatures above the first-order phase transition temperature
$T_c$\cite{note0}. The simulations were done at three different
couplings, and the lattice size in the Euclidean time direction,
$N_\tau$, was chosen to be 4. The temperature assignments at the
couplings $\beta=5.1$ ($T=1.5T_c$) and $\beta=5.15$ ($T=2T_c$) are
made using the results of previous simulations at $N_\tau=6$ and 8
\cite{nt46,nt8}. We also made a simulation at $\beta=5.35$, corresponding
to $T=(2.9\pm0.1)T_c$ \cite{note1}.  We have chosen the bare quark mass
to be $m=0.02T_c$.  We performed a finite size scaling study at $2T_c$
using $4\times16\times10^2$, $4\times24\times10^2$, $4\times24\times12^2$
and $4\times16^3$ lattices. At $1.5T_c$ we worked with a lattice size
of $4\times16\times10^2$ and at $2.9T_c$ with $4\times18^3$. The lattice
sizes have been chosen so that the correlators can be followed in at least
one direction to spatial distance of $2/T$ or more, and the transverse
size of the lattice is large enough to avoid spatial deconfinement. In
thermal equilibrium, the HMC trajectory lengths were taken to be one
unit of molecular dynamics time except for the cubic lattice at $2T_c$
where it was taken to be half an unit. Thermalisation was monitored using
four variables--- the spatial and temporal plaquettes, the Wilson line,
and the quark condensate.

We have analysed correlation functions constructed from Wilson loops of
various sizes (glueball-like correlators) for the first time in full QCD
at $T>0$. We have also analysed meson-like correlators, {\sl i.\ e.\/},
those constructed with a $\bar qq$ pair.  The particular operators that
we have examined belong to the scalar and vector irreps of the $T=0$
theory. Under the symmetry of the spatial transfer matrix for $T>0$,
{\sl i.e.\/}, the group called $D_4^h$, the former belong to an irrep
called the $A_1^+$. The vector irrep of $T=0$ reduces under the $T>0$
symmetry group: two linear combinations give the irrep $A_1^+$ and one
gives an irrep called $B_1^+$.

The glueball-like operators used in this study were the plaquette and
the planar 6-link loop. Noise reduction was done through a variant of
the fuzzing technique.  The operators were constructed at each of five
levels of fuzzing, and a variational method was used to extract the
lowest mass in the channel and project out the corresponding correlator.
More details of the methodology can be found in \cite{glue}.  We could
follow the correlation function in the $A_1^+$ sector up to a distance of
4, enabling us to obtain its mass with a reasonable degree of confidence.

The meson-like operators used here are local operators corresponding to
the $\pi$, $\sigma$, $\rho$ and $A$ at $T=0$. These screening correlators
have been studied for $T>0$ in the $A_1^+$ channel earlier \cite{meson}.
We report below our measurements in this channel.  The first measurement
in the $B_1^+$ channel was reported recently \cite{b1p}.  At all three
temperatures studied here, the $B_1^+$ correlators vanished within errors,
consistent with free field theory and unlike the $B_1^+$ correlators in
the glue sector obtained from the same simulations.

We made direct measurements of the autocorrelation function for the
$\pi$-like correlator, $C_\pi(z)$, at $T=2T_c$ on the
$4\times24\times10^2$ lattice. For the correlators at $z=0$ and $z=6$
we found comparable autocorrelation times, somewhat less than 1 MD
trajectory. For $z>6$ autocorrelations are hard to measure directly. We
limit them by the following argument.  For measurements of a correlator
$C(z)$ where the relative errors $e(z)/C(z)$ are expected to be similar
in magnitude, strong decreases signal large autocorrelations. Plotting
$e(z)/C(z)$ for the $\pi$-like correlator as a function of $z$, we
found no decrease for $z$ close to the center of the lattice. This
makes the existence of strong autocorrelations rather unlikely, and
eliminates possible simulation algorithm related systematic errors in
mass measurements.

In the meson-like channels we have estimated the screening masses by two
independent techniques--- fitting and constructing local masses. On all
lattices these two methods gave consistent results.

\begin{figure}[htbp]\begin{center}
   \leavevmode
   \psfig{figure=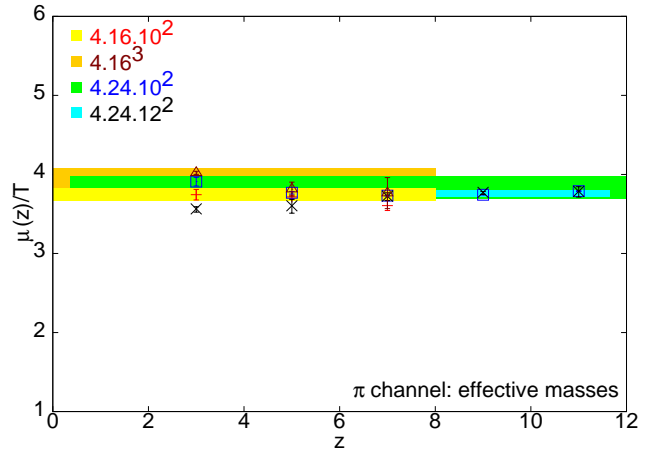,height=6cm,width=8.5cm}
   \end{center}
   \caption{The $A_1^+$ local mass obtained using local pionic operators at
      $T=2T_c$ on four different lattices. Also shown are the 1-$\sigma$ band
      on the corresponding fitted masses.}
\label{fg.fss}\end{figure}

Our finite size scaling study showed that finite volume effects are
under control. Figure \ref{fg.fss} shows the local masses obtained from
the pion-like correlator on the four different lattices at $2T_c$. Also
shown as bands are our estimates of the screening masses obtained by a
fit to the same correlator---
\beq
   \mu_\pi(A_1^+)/T = \cases{
          3.74\pm0.08 & ($4\times16\times10^2$ lattice),\cr
          3.92\pm0.16 & ($4\times16^3$ lattice),\cr
          3.83\pm0.14 & ($4\times24\times10^2$ lattice),\cr
          3.76\pm0.04 & ($4\times24\times12^2$ lattice).}
\eeq
Comparing the long ($N_z=24$) and short ($N_z=16$) lattices, we find
that the effective masses stabilise very well to a common value.
At the same time, this estimate of the screening mass is independent
of the transverse spatial size of the lattice, indicating that the low
lying eigenvalues of the transfer matrix have become independent of the
lattice size. Therefore we have determined the infinite volume screening
mass accurately.  Similar lattice size independence of screening masses
was observed in the other channels as well.

The screening masses from the $\pi$-like and $\sigma$-like correlators
were equal within errors, as were those from the $\rho$-like and the
$A$-like correlators. We investigated possible mass splittings more
carefully by making simultaneous fits to these pairs of correlators,
taking into account the covariance of the data due to the fact that the
correlators are constructed from the same configurations. This analysis
showed no hint of any mass splitting. The results are collected in
Table \ref{tb.masses}.

In principle, the $\sigma$-like correlator in the 4-dimensional theory
could mix with the vacuum and hence with the $A_1^+$ glueball-like
correlator. However the $\pi$-like channel does not mix with the
4-dimensional vacuum, and the above degeneracy of $\mu_\pi(A_1^+)$ and
$\mu_\sigma(A_1^+)$ cannot be explained if the latter is small due to
mixing with $\mu_g(A_1^+)$.  Hence screening masses seen in these two
channels are characteristic of fermion-bilinear currents in the theory.
Furthermore, our results taken in conjunction with earlier studies at
different cutoffs \cite{meson} indicate that $\pi$-$\sigma$ degeneracy
and the small value of their common screening mass, at these temperatures,
are cutoff independent.

Our best estimates of the $A_1^+$ projection of the $\rho$-like screening
mass is---
\beq
   \mu_\rho(A_1^+)/T = 5.6\pm0.1
\label{mrho}\eeq
at $T=2T_c$.  For $N_\tau=4$, free field theory gives
$\mu_\rho(A_1^+)/T=5.27$.  At other temperatures we also found values of
$\mu_\rho(A_1^+)/T$ close to, but slightly higher than, this perturbative
result.  The ratio $\mu_g(B_1^+)/T$ was found to be close to this,
but much larger statistics are needed to have any reliable estimate of
this number.

\begin{figure}[htbp]\begin{center}
   \leavevmode
   \psfig{figure=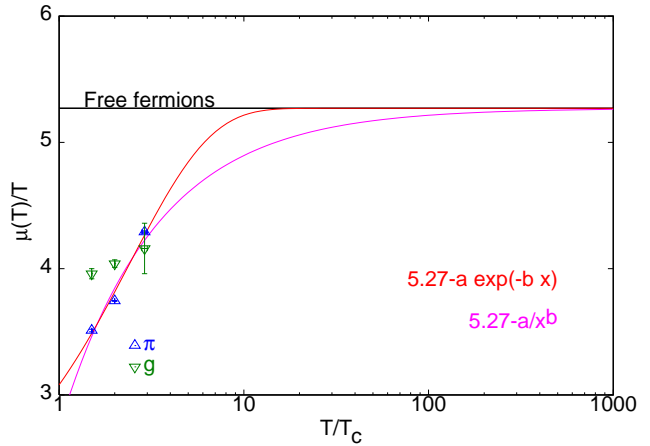,height=6cm,width=8.5cm}
   \end{center}
   \caption{The $A_1^+$ screening masses obtained using local pionic
       and glueball operators shown as a function of temperature along
       with phenomenological fits to the former. The fit formulae are
       in the same vertical order as the fit lines at large $T/T_c$.}
\label{fg.extr}\end{figure}

In Figure \ref{fg.extr} we show $\mu_\pi(A_1^+)/T$ and $\mu_g(A_1^+)/T$
as a function of $T/T_c$. In contrast to the slow variation of
$\mu_g(A_1^+)/T$ with changing $T$, we see that $\mu_\pi(A_1^+)/T$ is
increasing. The latter ratio is smaller up to $2T_c$, but there is
a cross-over near $2.9T_c$.  It is possible that at sufficiently large
$T$ the perturbative value $\mu_\pi(A_1^+)/T=\mu_\rho(A_1^+)/T=5.27$
is reached \cite{note2}. In accordance with this, we tried various
smooth and monotonic two parameter fits to the data which have this
limit as $T\to\infty$, and have a finite and non-zero value at $T=T_c$.
Two such fits are shown in the figure. Extrapolating these to higher
temperatures, we find that $\mu_\pi(A_1^+)$ becomes compatible with
free fermion field theory only for $T_p\approx10T_c$ or more. It would be
interesting to test this extrapolation by direct simulation \cite{note3}.

Our results have implications for the dimensionally reduced theory. If
the perturbative approach to dimensional reduction were accurate at
these temperatures, then one should be able to integrate out the
Fermions altogether. Their contribution would be visible only in the
effective couplings between the zero momentum gauge modes. At the same
time, $\mu(A_1^+)$ should be independent of whether it was measured in
the $\pi$-like or $\rho$-like channel and, being close to the
free-field value, should decouple from the long distance physics.
However, something non-perturbative happens in the Fermion sector at
experimentally accessible temperatures of up to about 1 GeV.

Since $\mu_\pi(A_1^+)<\mu_g(A_1^+)$ up to $3T_c$, this excitation cannot
be integrated out of the effective theory that describes physics at
the length scale $1/\mu_g(A_1^+)$.  This state of affairs cannot be
accommodated into a perturbative description of dimensional reduction,
since all Fermions and multi-Fermion states need to be integrated out in
such an approach.  For $3T_c<T<10T_c$, $\mu_\pi(A_1^+)$ still deviates
sufficiently from its perturbative value to obstruct perturbative
approaches to dimensional reduction, although the obstruction becomes
progressively smaller.

The values of $\mu_g(A_1^+)/T$ we measured in 4-flavour QCD are 53\%
higher than in quenched QCD \cite{glue}. Although comparisons of
glueball masses at $T=0$ in quenched and full QCD show much smaller
shifts \cite{wuppertal}, the increase in the Debye screening mass, in
going to the four-flavour theory is about 30\% in leading order of the
perturbation expansion at any $T$.  At the same order, the increase in
$\mu_g(A_1^+)/T$ is expected to be about 60\%. Although perturbation
theory may not be reliable here, this estimate shows that the large
shift is not unnatural.  In contrast, $\mu_\pi(A_1^+)/T$ decreases
by 17\% when going from quenched to 4-flavour QCD.  The obstruction
to dimensional reduction appears very clearly in 4-flavour QCD due to
these two movements. However it is present even in quenched QCD, albeit
in a subtle form, since $\mu_g(A_1^+)<\mu_\pi(A_1^+)<\mu_g(B_1^+)$
for $T\approx1.5$--$2T_c$.

This first measurement of $\mu_g(A_1^+)$ in full QCD thus sheds new light
on the effective long distance theory at $T>T_c$.
As far as gauge invariant correlators are concerned, the situation up to
$2T_c$ is reminiscent less of the usual dimensional reduction than of the
$T=0$ situation where the long distance effective theory is built from
scalar and pseudo-scalar fermion bilinear composites.  Thermodynamics,
governed by $\Lambda_0(T)$ alone, cannot be explained by this effective
model.  In any case, there are differences between the $T=0$ and $T>0$
effective theories. At $T=0$ chiral symmetry is broken and the scalar
and pseudo-scalar masses are different. Here, on the other hand, this
symmetry is restored and the masses are equal.

Finally we mention that a weaker form of dimensional reduction may still
be valid in this range of temperatures. This can be investigated by
looking for degeneracies of all measurable screening masses and checking
whether they reflect the 4-dimensional $T>0$ symmetries or an effective
3-dimensional symmetry. This requires a measurement of correlations of
non-local Fermion bilinear operators, and is left for the future. Also
left for the future is a detailed investigation of the continuum limit
of our results.  A similar investigation in the more realistic case
of 2 light and one medium heavy flavour of quarks is a desirable future
computation, specially with Wilson Fermions.

\begin{table}[htbp]\begin{center}
  \begin{tabular}{ccccc}  \hline
  $T$ & statistics & $\Delta \mu/T$ & $\mu_\pi(A_1^+)/T$ & $\mu_g(A_1^+)/T$ \\
  \hline
  $1.5$ &  900 & $0.04\pm0.14$ & $3.51\pm0.01$   & $3.96\pm0.04$ \\
  $2.0$ & 1875 & $0.06\pm0.24$ & $3.744\pm0.001$ & $4.04\pm0.03$ \\
  $2.9$ &  825 & $0.03\pm0.06$ & $4.29\pm0.01$   & $4.16\pm0.20$ \\
  \hline
  \end{tabular}\end{center}
  \caption[dummy]{Statistics and measurements of various screening masses and
     the splitting $\Delta \mu\equiv \mu_\pi(A_1^+)-\mu_\sigma(A_1^+)$ at
     different $T$. For $T=2T_c$ the results come from the $4\times24\times10^2$
     lattice.}
\label{tb.masses}\end{table}

\end{multicols}

\begin{thebibliography}{99}
\bibitem{note1}
  From \protect{\cite{mtcmass}} we know that the ratio of the lattice
  spacings, as determined by the $T=0$ $\rho$ mass, at $\beta=5.15$ and
  5.35 is $1.45\pm0.05$. The same result is also obtained by setting the
  scale through the proton mass.
\bibitem{note2}
  In free field theory we have $\mu_\pi(A_1^+)=\mu_\rho(A_1^+)$. However,
  since $\mu_\rho(A_1^+)$ is shifted a little, indicating small interaction
  effects, the same effects could induce a splitting between these two masses.
  On the other hand, at high temperatures, if the large effects in $\mu_\pi$
  do go away, then this small shift in $\mu_\rho$ could also become smaller.
  For this reason we have taken the asymptotic value of $\mu_\pi(A_1^+)$ to be
  the free field value. However, no significant change in $T_p$ occurs with
  the choice $\mu_\pi(A_1^+)/T=5.6$.
\bibitem{note3}
  Extending our computations to higher temperatures is not simple. Finding
  $\beta$ corresponding to $T\approx10T_c$ needs a sequence either of 
  direct computations of the critical couplings at larger $N_\tau$, or
  of $T=0$ computations to determine lattice spacings.
\end{thebibliography}
\end{document}